\documentclass[11pt]{article}
\usepackage{a4}
\usepackage{times}
\usepackage{graphicx}
\usepackage{color}
\usepackage{subfig}
\usepackage{multirow} 

\begin{document}
\title{Measurement errors and scaling relations in astrophysics: a review}
\author{
S. Andreon,$^1$\thanks{stefano.andreon@brera.inaf.it}, M. A. Hurn,$^2$
\\
$^1$INAF--Osservatorio Astronomico di Brera, Milano, Italy\\
$^2$University of Bath, Department of Mathematical Sciences, Bath, UK\\
}

\maketitle

\begin{abstract}
\noindent
This review article considers some of the most common
methods used in astronomy for regressing one quantity against another
in order to estimate the model parameters or
to predict an observationally expensive quantity using trends between
object values.
These methods have to tackle some of the awkward features prevalent in astronomical
data, namely heteroscedastic (point-dependent) 
errors, intrinsic scatter, non-ignorable data collection and 
selection effects, data structure and non-uniform population (often
called Malmquist bias), non-Gaussian data, outliers and mixtures
of regressions.
We outline how least square fits, weighted least squares methods, Maximum Likelihood, survival analysis,
and Bayesian methods have been applied in the astrophysics literature when one or more of these features is present.
In particular we concentrate on errors-in-variables regression and we advocate
Bayesian techniques.
\end{abstract}

\section{Introduction}

In astrophysics, the term ``scaling relation'' is taken to mean a relationship
between different physical values of astronomical objects such as, stars, 
galaxies, galaxy clusters and black holes. 
In more statistical terms, the term indicates a relationship between two variables, Equation~(\ref{eqn:1}), along with an associated measure of variability associated with that relationship.
Some well-known examples are the
Tully-Fisher relation (between galaxy luminosity and circular velocity), the
Faber-Jackson relation (between galaxy luminosity and velocity dispersion), the
fundamental plane (between galaxy brightness and velocity dispersion and size),
the Magorrian relation (between the mass of a black hole and the 
bulge mass). These relationships are of considerable interest
because from their existence and from their parameters we can elaborate
or test theories about
the object formation and evolution. At a high level, we can make inference from
these relationships about the Universe's birth and fate, or the mysterious dark energy, etc.
For example, the Cepheid period-luminosity scaling informs us about
the physics regulating these objects, and, at the same time, in the 1930's
demonstrated that the Universe is larger than
our own Galaxy, that nebulae are outside our own Galaxy and that the Universe
is formed by galaxy-islands (\cite{Hubble25}).

Unfortunately it is rarely possible to measure values completely without
error. Errors may be deterministic (e.g. rounding to the nearest integer), stochastic 
or both. In this
article, we concentrate on the problem of stochastic noise and we review the
interesting statistical questions which arise and the associated methodology
currently in the literature. We begin with a simple example to illustrate some
of the points which may occur. Suppose we are given N pairs of observations,
$x^{obs}_{i}$ and corresponding $y^{obs}_{i}$ for data points $i=1,...,N$.  
The $x^{obs}_{i}$ are observations on the true values $x_i$, 
while the $y^{obs}_{i}$ are observations on the true values $y_i$. It would
be common in this field to plot $y^{obs}$ against $x^{obs}$ to  try to learn
something either about the objects under study, or, via them, about a bigger
objective, such as the geometry of our Universe. Expressed more
formally, estimating 
the relationship between the
the true values
$x_i$ and $y_i$
is of interest
\begin{equation}
y_i = f(x_i, \theta) 
\label{eqn:1}
\end{equation}
where $\theta$ represents any parameters of the relationship.
In the simplest case, the function $f$ could be a linear relationship, 
with $\theta$ representing the slope $a$ and intercept $b$:
\begin{equation}
y_{i} = a x_i +b .
\label{eqn:2}
\end{equation}
If 
the $\{x_i\}$ are actually observed without error 
(in other words, $x_i^{obs}=x_i$) and the $\{y_i\}$ are
each corrupted by an additive Normal, or Gaussian, error with constant variance,
$y_i^{obs} = y_i + \epsilon_i$ where $\epsilon_i \sim N(0,\sigma^2)$, for
$i=1,\ldots,N$, then this problem is the familiar linear regression (the notation $\sim$ reads ``is distributed as" throughout).
Sadly for astronomers, but perhaps happily for statisticians, things are rarely this simple.

In this article, we will begin in Section~\ref{sec:regress} by discussing some
of the reasons why astronomers are interested in regression. Section~\ref{sec:features} describes some of the complicating
features that are common in astronomical data. In light of these features,  we
review the current state of the literature, mostly astronomical, in Section~\ref{sec:review}. 
Section~\ref{sec:perform} will compare the performances of some of these regression methods
on simple data. 
This
review only addresses scaling relations in astronomy between physical
values,  given the breadth of the applications and the volume of
statistical methodology involved we do not attempt to tackle spatial or
temporal models.

\section{Why astronomers regress}
\label{sec:regress}

There are a number of different reasons why astronomers may have an interest in 
regressing one quantity against another.
These include

\begin{figure}
\centerline{\includegraphics{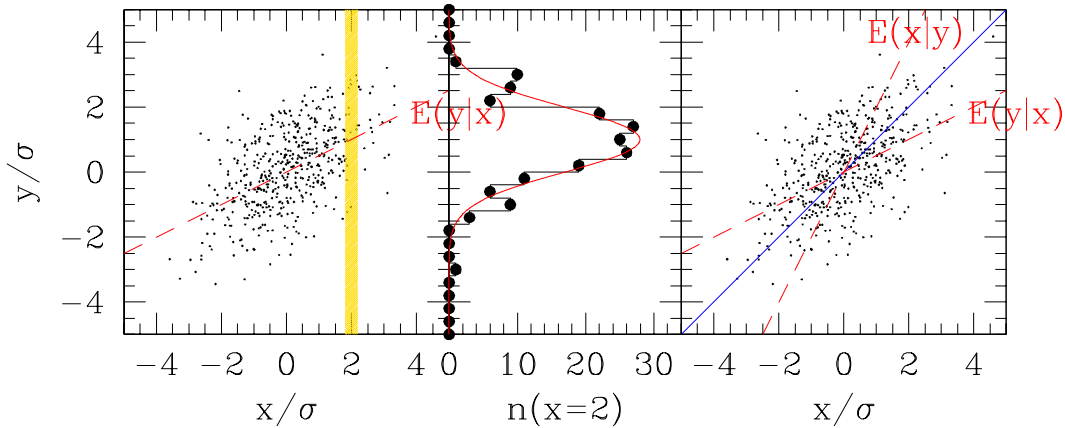} }
\caption{Left panel: 500 points drawn from a bivariate Gaussian
overlaid by the line showing the expected value of $y$ given $x$;
the yellow band covers those $y$ for which $x$ is
close to 2.
Central panel: Distribution of the $y$ values for $x$ values in a narrow band of $x$ centred on $2$, as shaded in the left panel. 
Right panel: as the left panel, but adding the lines giving the
expected $x$ values at a given $y$, and the $x=y$ line.
\label{regr.dirinv}}
\end{figure}

\begin{itemize}

\item {\it Parameter estimation.}
It may be that the parameters of the relationship between $x$ and $y$ are
themselves the primary interest, as in the slope of the cluster of galaxies
$L_X-T$ (X-ray luminosity vs Temperature) scaling relation. 
Usually, we are interested in both the parameter estimates and their estimated
uncertainty. When the relationship is important because of what it implies for a
higher-level question (for example, Universe geometry), then it is particularly
important to carry forward the parameter uncertainty into any higher-level
inference. We emphasise that, most of times, we are interested in 
the relation between the true values $x$ and $y$, rather than between the observed
values
$x^{obs}$ and $y^{obs}$.

\item {\it Prediction.} It may be the case that there is a considerable
difference in either the difficulty or the cost of measuring $x^{obs}$ and
$y^{obs}$. Provided there is a trend between the two values and not too much
scatter, it might then be desirable to predict the expensive values of $y$ (say)
using the cheaper measurements of $x$, $x^{obs}$. Two examples from astronomy are the use
of colours to predict redshift (see \cite{koo85} for example),  and the use of one of
the many cheaper values to predict mass (for example richness
\cite{andreon2010scaling}). The emphasis here is on assessing the uncertainty
associated with the predicted values of $y$.

\item {\it Model selection.} It may be that 
the form of the relationship between $x$ and $y$ is of primary interest, 
that is determining the particular $f(x,\theta)$ in Equation~(\ref{eqn:1}), with all this implies for the astrophysics.
The important topic of more formal model choice
is outside the scope of
this review, see instead for example \cite{liddle2007information}
and \cite{trotta2008bayes}.

\end{itemize}
To illustrate how the goals of an analysis can alter 
what may be considered a good model,
Figure~\ref{regr.dirinv} shows a set of 500 points drawn from a
bivariate Gaussian where marginally both $x$ and $y$ are
standard Gaussians with mean 0 and variance 1 and $x$ and $y$
have correlation $1/2$.
What is the ``best'' line explaining the relationship between $x$ and $y$ here?
If we are trying to summarise the relationship itself, then 
the blue $y=x$ line shown on the right hand panel seems reasonable.
But is this also good if instead we were trying to predict a new value of $y$ given a value of $x$?
Superimposed in red on the left hand panel of Figure~\ref{regr.dirinv} is the
line giving the theoretical conditional expectation of
$y$ given $x$ (known theoretically for this bivariate
Gaussian to be $y = x/2$).
Why might this be a better line for the purposes of prediction?
Consider a small range of $x$ values close to 2 indicated in yellow.
The middle panel indicates the corresponding $y$ values of the points
captured in this way (overlaid on the theoretical distribution of $y | x=2$, where the symbol $|$ will be used throughout to indicate ``conditional on").
The average of the $y$ values falling in the shaded area is closer
to the value predicted by the red line (a value of 1 in this case) than to
the value predicted by the blue, $y=x$, line (2 in this case).
Although this line then is good for prediction, it perhaps seems too shallow with respect to 
the overall trend identified by the points, which is better
captured by the $x=y$ blue line.
\cite{isobe1990linear} provides a nice discussion of these issues, and
we offer a further example in Section~\ref{sec:perform}.

Figure~\ref{regr.dirinv}, taken from 
\cite{andreon2010scaling}, illustrates one further point, namely the asymmetry of regression.
If, rather than predicting $y$ from $x$, one wishes to do the reverse and predict $x$ given $y$, then the line defined by the conditional expectation of $x$ given $y$ may be useful.
Shown in red in the third panel, this line is different again from the previous two.
In this example, the symmetry of the marginal distributions of $x$ and $y$ is reflected in the symmetry of the two red lines.
However there is an underlying conceptual difference between the two
predictions given that we regard one variable as a predictor and the other as a response.
In a slightly more complex setting than this, \cite{lauer2007selection} emphasises
that while this is known to many astronomers, not all appreciate
the consequences of the difference between the direct and inverse
fit in the context of the Magorrian relation (where 
$x=L$ is the galaxy luminosity and $y=M_\bullet$ is the black hole
mass).
Astronomers may often find themselves in situations where a symmetric treatment 
of variables would be more natural than a ``predictor-response'' approach.

\begin{figure}
\centering
\subfloat[]{\includegraphics[width=0.45\textwidth]{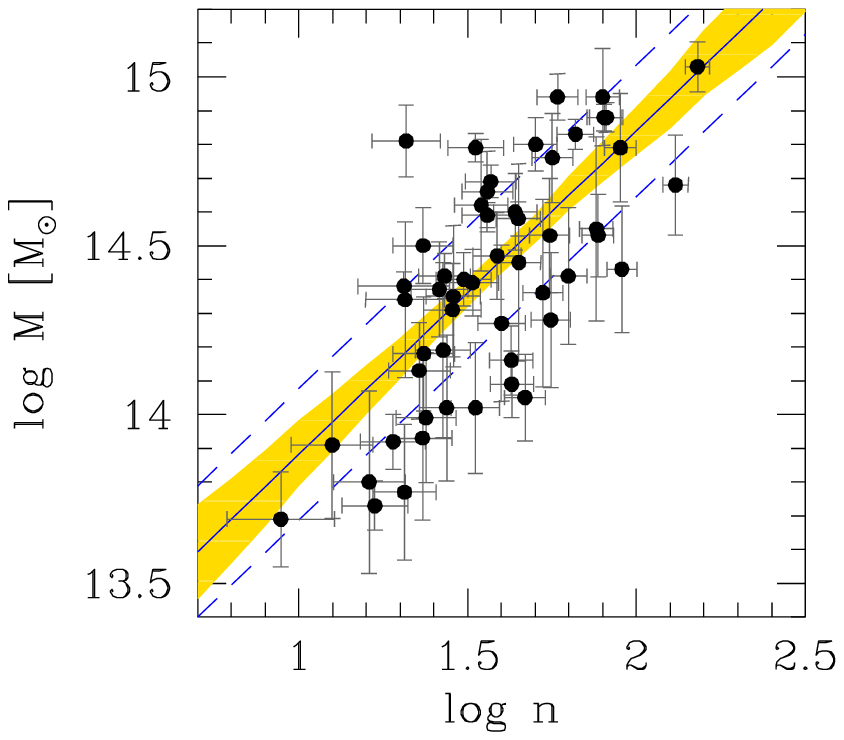}}
\qquad
\subfloat[]{\includegraphics[width=0.45\textwidth]{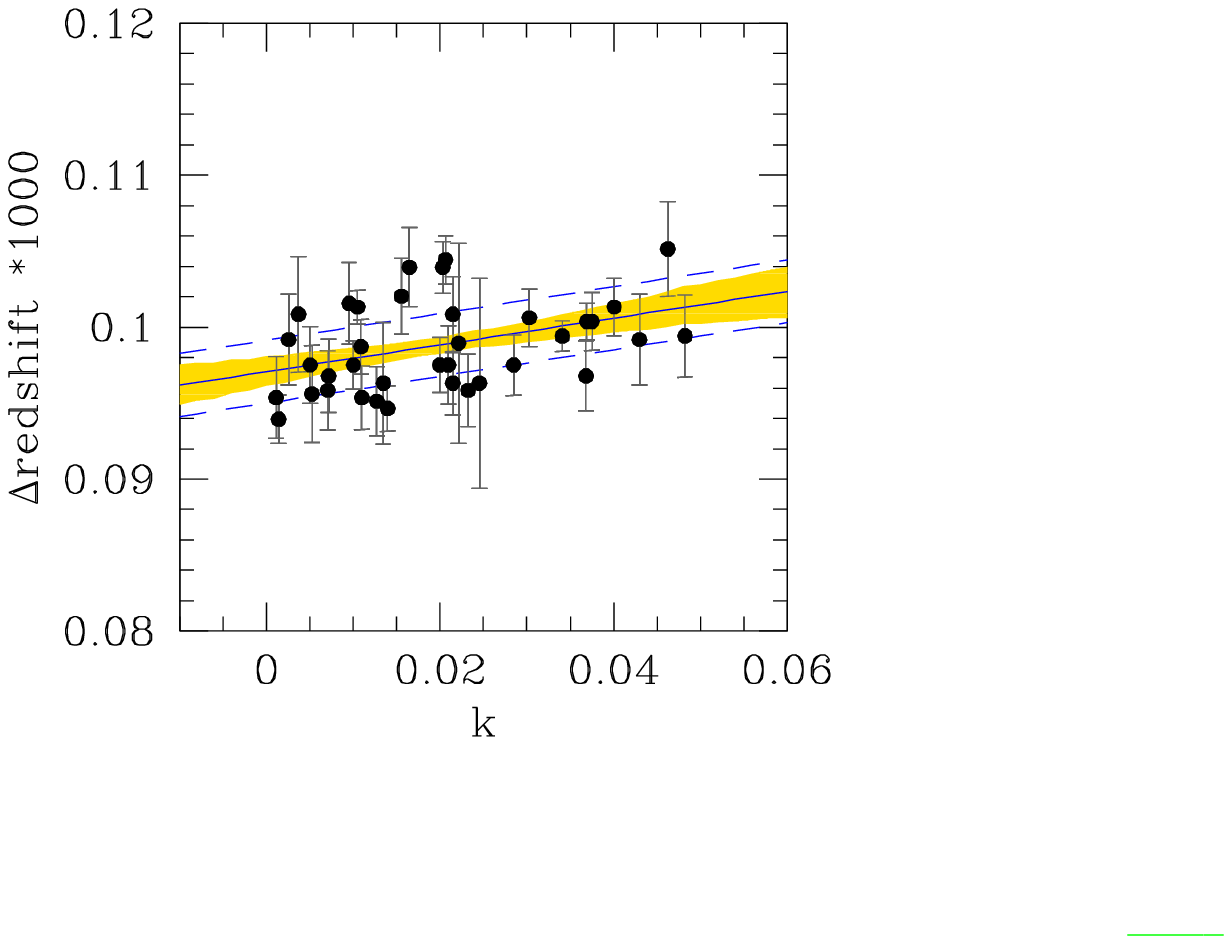}}
\caption{
{\it (a)}
Mass ($M$) vs Richness ($n$) 
(taken from \cite{andreon2010scaling});
these data represent galaxy clusters, each point corresponding to
a different cluster.
Mass is one of holy grails in 
astronomy but it is observationally expensive or even impossible, to measure, whilst  
at the same time being needed in many studies because
most properties depend on mass. The easily observed quantity
richness ($n$) could in this respect be useful to
predict mass.
{\it (b)}
Wavelength shift ($1000*(\Delta \lambda / \lambda - 3.0248)$), vs sensitivity ($k$) 
to physical constant (data taken
from \cite{ubachs2004highly});
the points correspond to different 
measurements on a single object, the quasar 3C295.
One of the pressing questions in astronomy and physics is whether
physical constants are indeed constant or take different values at different
locations in the Universe or at different times.
The slope of the regression, if different from zero, 
implies that (a specific combination of) physical constants are not 
taking a unique value everywhere in the Universe, being different
between Earth laboratories and the very distant Universe probed
by the plotted experimental points. 
In both panels
the solid line marks the fitted regression, taking into account the intrinsic 
scatter as well as the measurement noise, while the 
dashed lines
show this plus or minus the estimated intrinsic scatter $\sigma_{scat}$ derived from
the Bayesian analysis in \cite{andreon2010scaling}.
Error bars on the data points represent the scale (size) of measurement errors, 
indicating one standard deviation.
Distances between the data and the regression are due to both 
measurement error and intrinsic scatter.
(The shaded regions mark 68\% highest posterior credible intervals for the regression.)
} 
\label{fig.scatter}
\end{figure}

\section{Common features of astronomical data}
\label{sec:features}

\subsection{Heteroscedastic error structure}

Much standard statistical theory deals with the situation where all
observations are made with the same accuracy in the sense that the sizes (variances) of the
errors are assumed equal, say $\sigma^2_y$ for all the observations, e.g
\begin{equation}
y^{obs}_i = y_i + \epsilon_i \ \ \mbox{where } \epsilon_i \sim N(0,\sigma^2_y),\ i=1,\ldots, N .
\label{eqn:homo}
\end{equation}
Such a simplifying assumption is unlikely to be valid in many astronomical studies where a heteroscedastic error structure may have to be adopted, e.g replacing Equation~(\ref{eqn:homo}) by
\begin{equation}
y^{obs}_i = y_i + \epsilon_i \ \ \mbox{where } \epsilon_i \sim N(0,\sigma_{y,i}^2),\ i=1,\ldots, N .
\label{eqn:hetero}
\end{equation}
The notation $\sigma_{y,i}^2$ is adopted here for later compatibility with situations where the related variable $x^{obs}_i$ also exhibits heteroscedastic error.
There are a number of reasons why Equation~(\ref{eqn:hetero}) may be more
appropriate, sometimes because the $N$ objects may not be observed under
identical conditions, for example, under variable atmospheric conditions,
sometimes because of object related differences,  for example other values
of the object such as luminosity, colour, size, etc. As an additional
complication, it is possible for the error structure on the $i^{th}$ object,
$x_i$ and $y_i$, to be correlated; \cite{akritas1996linear} discusses the
example of the Tully-Fisher relation where an inclinational correction applied
to a galaxy affects both its luminosity and velocity errors.

\subsection{Intrinsic scatter}

Equations~(\ref{eqn:1}) and (\ref{eqn:2}) assume that the relationship
between the $N$ objects' values $\{x_i\}$ and $\{y_i\}$ is deterministic, albeit unknown.
In many cases, this clean relationship is not representative and there is also an intrinsic scatter (a stochastic variability) to consider.
This extra layer of variability is distinct from the (possibly heteroscedastic) measurement noise,
indicating instead that the objects under study are 
examples drawn from a population with a spread of values.
Extending the model in Equation~(\ref{eqn:2}) to include intrinsic scatter, the full set of equations becomes
\begin{eqnarray}
x_i^{obs} & = & x_i + \epsilon_{x,i} \mbox{ where } \epsilon_{x,i} \sim N( 0 , \sigma^2_{x,i} ) \nonumber \\
y_i^{obs} & = & y_i + \epsilon_{y,i} \mbox{ where } \epsilon_{y,i} \sim N( 0 , \sigma^2_{y,i} ) \nonumber \\
y_{i}  & \sim & N( a x_i +b , \sigma^2_{scat} )  .
\label{eqn:full}
\end{eqnarray}
At this point a formal distinction between the two types of measurement error found in the literature should be noted
since we have made a choice by specifying Equation~(\ref{eqn:full}): the situation above where a distributional
assumption is made about the observed data given the true values is called the ``classical measurement error'' model;
if the reverse is true and the distributional assumptions are made about the true values given the observed ones,
then the model is known as ``Berkson error'' (\cite{buonaccorsi2009measurement}, Chapter 1).   We will work
throughout with the classical model, because the latter is usually available in physics and astronomy
and called calibration when dealing with instruments: one injects $x$ in the apparatus and measure the distribution of
observed data, $p(x^{obs}|x)$.

Two examples where intrinsic scatter is relevant are shown in Figure~\ref{fig.scatter}.
The left panel of Figure~\ref{fig.scatter} shows a mass vs richness scaling where the variability between objects is greater than that simply due to measurement errors and is more likely due to differences of the physics on the objects under study (\cite{andreon2010scaling}).
In other cases, the scatter may be due to unaccounted systematic effects.
The right hand plot reports the ratio between the wavelength of several spectral lines compared to the
wavelength measured in laboratory,
$(\lambda_{obs}-\lambda_{lab})/\lambda_{lab}=\Delta \lambda / \lambda$
against the parameter $k$, measuring the sensitivity of the considered
spectral lines to the physical constant $k$ (\cite{ubachs2004highly}).
The intrinsic spread visible is due
to a source of error still not identified, because all these measurements pertain
to a single object and so there is no ``population'' variability effect.

\subsection{Non-ignorable data collection and selection effects}

In an ideal world, the sample we have is randomly selected from the population we wish to study,
and so any inference we draw about the sample  applies to the population; in other words, we can ignore the way
in which the data were collected. Unfortunately, data collection is very often non-ignorable. As
an example, in \cite{lauer2007selection} studying black hole masses, galaxies at high redshift
are selected by their nuclear activity while galaxies at low redshift are selected by luminosity
or velocity dispersion. But as that paper points out, these two subpopulations are  not
homogeneous because of the spread in luminosity at any given black hole mass and therefore the effect
of the selection criteria may be different for the two. In this particular case, ignoring this
selection effect difference leads to biased estimation.
At this point, it may be worth giving a formal definition of bias.
Statisticians define the ``bias'' of estimators to be the difference between the expected value of the estimators and the true value of the parameters; however the term ``biased'' is also widely used to refer to the non-representative nature of a sample (effectively the source of the statistical bias).

\begin{figure} 
\centerline{\includegraphics[scale=0.8]{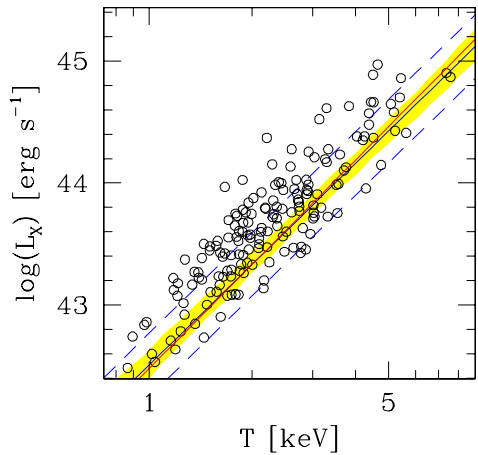} }
\caption{A simulation (from \cite{AandH11}) built around a true $L_X-T$ data set and selection function
(from \cite{pacaud2007xmm}), illustrating how the observed
data may be unrepresentative compared to the underlying scaling relation due
to selection effects. The solid blue line shows the true relationship but due to
the selection effects, most points lie above the line.}
\label{fig.LxT}
\end{figure}

Selection effects can arise in a number of forms:
a) Object values may be partially missing in a non-random way although the object is in the catalogue.
This means we can say something about the missing object, e.g. its position
in the sky, but we are lacking a measurement of the quantity of interest; 
b) The object itself is missing from the catalogue (not at random), and often
we can say little or nothing about what is missing, including any idea of
how many objects are missing;
c) In an extension of the above scenarios, objects can be missing with a probability
different from zero or one, for example objects above/below some threshold enter/exit the sample with some probability $p$, where $p$ is usually a smooth
function of the object's parameters. This case is illustrated in in Figure~\ref{fig.LxT}, taken
from \cite{AandH11}. At all
temperatures $T$, clusters brighter than average are easier to detect (simply because
they are brighter) and therefore they are over-represented in most samples, while those
fainter than the average are underrepresented. Therefore, at a given $T$
the mean $L_X$ of clusters in the sample is higher than the average $L_X$ of the
population at the same $T$. The selection is stochastic:
it is possible to miss an object above
the threshold and yet still to record some objects below the threshold.
This effect is self-evident and is called the ``bias''
of the X-ray selected clusters in \cite{AandM11}, and the ``halo
model'' in \cite{allen2011cosmological}, but its consequences are not always fully 
appreciated, as emphasised in \cite{AandM11}.
We will return to these ideas in Sections~\ref{sec:asurv} and
\ref{sec:bayes.full}.

Since collecting samples with known (and thus hopefully correctable) selection 
functions is observationally hard, the issue of selection effects 
and modelling data collection
is often deferred until a selection effect becomes apparent in the data
themselves. 
Often, astronomers
are forced to study the property of the population entering in the sample
(SDSS or Galex galaxies, PG QSO, EROs), instead of the property of
the population sample, because either the selection function is poorly
known or it is so severe that it is impossible to make inference about the wider population.

\subsection{Data structure and non-uniform populations}

Even when the data collection does not contain selection problems, neglecting the population structure can induce estimation biases: 
\cite{eddington1913formula} and \cite{malmquist1922some} reminded the astronomical community
that even symmetric errors (for example, Gaussian likelihoods) have an effect, later
called the Eddington or Malmquist bias.

\begin{figure} 
\centerline{\includegraphics[width=6truecm]{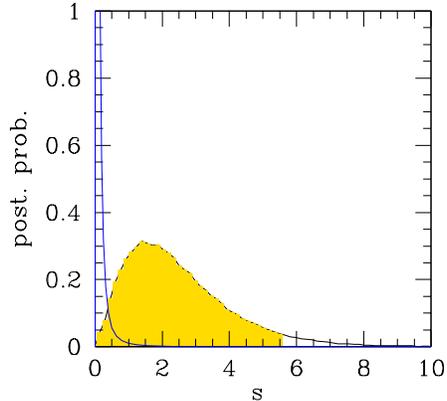} }
\caption{Posterior distribution (black curve) of the source flux, having observed four 
photons and knowing that source number counts have an Euclidean distribution, 
that is their probability distributions follow a power law
(shown by the blue curve).
The highest posterior 95\% interval
of the source flux is shaded in yellow (adapted from \cite{andreon2010bayesian}).
}
\label{fig.malmquist}
\end{figure}

Figure~\ref{fig.malmquist} illustrates
the problem using an example from the X-Bootes survey
(\cite{kenter2005xbootes}) of X-ray sources
performed with the Chandra Space Telescope.
The observation here $s^{obs}$ is a count measurement on the 
underlying flux, $s$.
As such, we can assume that given the underlying flux $s$, the data point $s^{obs}$ is distributed as a
Poisson random variable,
$s^{obs} \sim \mathcal{P}(s)$.
However it is also known from previous accurate observations, that the
probability distribution of $s$ (known as the number counts
by astronomers) 
is a power law with slope $-2.5$ (in the standard
astronomical units this slope is referred to as the Euclidean 
slope).
This distribution is illustrated in blue on Figure~\ref{fig.malmquist}
showing that small values of $s$ are more likely than large ones.
Therefore,
a source with $s^{obs}$ is more likely overall to be a fainter source with an
overestimated flux than a stronger one which has been underestimated.
This effect has nothing to do with the asymmetric Poisson 
distribution, it would still be there if a Gaussian noise structure were assumed.
Using Bayesian methodology (as described in Section~\ref{sec:bayes.full}),
Figure~\ref{fig.malmquist} 
shows the number counts and the posterior
distribution of the object flux for
a source with four observed photons, $s^{obs}=4$, typical of the 
X-Boote survey. The posterior
distribution has its mode at $s=1.5$ and mean at $s=2.5$, in
agreement with astronomical experience that the true flux, $s$, of
a source is likely to be lower than the observed value, $s^{obs}$.

Whether or not a Bayesian approach is preferred, most astronomers would agree that
some form of bias correction is often needed. Such bias is widespread in astronomy,
applying not only to fluxes, but almost to all quantities which have an abundance
gradient, such as galaxy cluster richnesses (number of galaxies) or masses,
parallaxes (i.e. the small angular displacement of stars due to the Earth's orbit
around the Sun),  velocity dispersions (how fast galaxies move inside clusters of
galaxies or how fast stars move in globular clusters). This has an obvious impact on
the determination of regression parameters of scaling relations using these
quantities, but its consequences are not always fully 
appreciated, as emphasised in \cite{andreon2010scaling}.

\subsection{Non Gaussian data, outliers and other complications}

Astronomical data, like all data, can be affected by a number of other problems which
negatively affect a regression analysis. For start, it is quite common for the error structure
to be other than Gaussian, particularly in situations where small numbers of counts are
recorded (even for large numbers of counts, a Gaussian approximation may be not
acceptable, see \cite{Humphreyetal09}).
For example, in the left-hand panel of
Figure~\ref{fig.scatter} the x-axis plots the log of a quantity which is 
(close to being) Poisson distributed.

Another very common problem is that of outliers and contaminated samples; as soon as
one collects a sizable sample, it is quite possible to also collect data for objects
having little to do with the population of interest.  When these objects fall far
away from the model suggested by the regression, they are called outliers and some
sort of outlier identification and removal may be needed, although removal may only
be essential in cases where the outlier is influential (see \cite{atkinson1985plots}, Chapters 1 to 5, for
more information on assessing outliers as well as other important diagnostic aspects
of regression modelling).

\begin{figure}
\centerline{\includegraphics[scale=0.6]{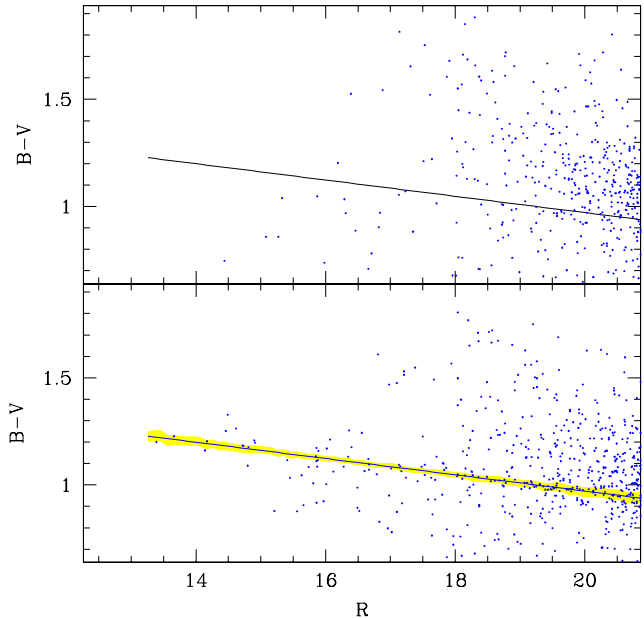} }
\caption{The $B-V$ colour vs $R$ magnitude 
relation of galaxies (from \cite{andreon2006new}), where $B-V$, 
short for blue minus visual magnitudes, is 
a colour index used in classification and the $R$ magnitude is $-2.5$ times 
the base 10 log of the source flux plus a constant. 
The spread in colour around the mean regression, called in astronomy the colour-magnitude relation, 
is a measure of heterogeneity in the star-formation histories
of red cluster galaxies; the larger the spread, the more variable their
star formation histories are.  
{\it Top panel:} diagram for galaxies in a random line of sight,
showing a realization of contaminating galaxies.
{\it Bottom panel:} diagram for interesting galaxies belonging to
the cluster Abell 1185, plus contaminating galaxies, i.e. galaxies in
the cluster line of sight but unrelated to the cluster.
Which individual point is interesting rather than
contaminating is initially unknown.
As a result interesting galaxies form a regression heavily contaminated
by outliers (there are 4 contaminants per interesting point).
The solid line is the mean colour-magnitude relation of 
interesting galaxies  fitted from the lower plot and overlaid on both plots, the narrow yellow area
marks the 68\% highest posterior interval (lower plot only).
}
\label{fig.A1185cm}
\end{figure}

There can be situations where the level of contamination by  data points not following the
hypothesised regression model makes it hard to identify the true population under study.
Figure~\ref{fig.A1185cm} gives an example of such a situation where the sample of interest is
overwhelmed by other points. In this case more powerful methods are needed to identify and remove
the contaminating sample (\cite{andreon2006new} and \cite{A06}). 

The sample could also contain several subpopulations, each following their own regression line
(indeed this is one way of viewing large scale contamination such as that in
Figure~\ref{fig.A1185cm}, with points either being in the group of interest or in the
contaminating subpopulation). Such a mixture of regressions greatly complicates the fitting
process (see \cite{hurn2003estimating} for examples of such mixtures of regressions in
non-astronomical examples and \cite{A06} for an astronomical application).

\section{Commonly used regression techniques in astronomy}
\label{sec:review}

In this section we will summarise the various approaches to regression  commonly used in
astronomy. 
Generally in cases where parametric models are to be used, it is possible to label approaches as either ``frequentist'' or ``Bayesian''.
The former bases inference on the likelihood of how the data arise, while the later 
works with the posterior distribution of the parameters given the data using some prior information in the model building.
Consequently, the rationale behind parameter estimation differs between the two, as does the subsequent assessment of parameter uncertainty.
In a frequentist setting, Maximum Likelihood Estimation (MLE) is usually preferred, that is choosing the parameter estimates which maximise the likelihood.
In a Bayesian setting, parameter estimates are chosen as appropriate summaries of the posterior distribution, for example the posterior mean or the posterior mode.
Other parameter estimation rationales are also in use, 
for example Method of Moments, Least Squares or Robust estimates; 
a recent overview is given by \cite{gillard2010overview}.
We review how some of this methodology have been used in the astronomy literature, attempting to highlight strengths and weaknesses. 

\subsection{Ordinary Least Squares regression}

The method of fitting a regression line $y=ax+b$ by ordinary least squares regression (OLS) is
extremely widespread in many disciplines (\cite{draper1998applied}). Based on the rationale of
minimising the sum of squared residuals (i.e. the difference between the observed data and the
values fitted to each point by the model), $S = \sum^N_{i=1}(y^{obs}_i-(b+ a x_i))^2$, 
the parameter
estimates are easily calculated as 
\begin{eqnarray}
\hat{a} & = & \frac{\widehat{\mathsf{Cov}}(x,y^{obs})}{\widehat{\mathsf{Var}}(x)} \nonumber \\
\hat{b} & = & \bar{y}^{obs} - \hat{a} \bar{ x} 
\label{eqn:OLS}
\end{eqnarray}
where we deliberately express the estimators in terms of the estimated moments 
of $x$ and $y^{obs}$ for compatibility with later estimators.

As a rationale for parameter estimation, OLS does not require any underlying distributional assumptions to be made.
Finding uncertainty estimates associated with these parameter estimates does however require some such
assumptions (as a minimum the mean and variance of the error structure for $y^{obs}$).
In the case that the $\{x_i\}$ are observed without any error and $y_i$ errors are all Gaussian, of the same size, and there is no intrinsic scatter 
(i.e. that the rather simplistic Equation~(\ref{eqn:homo}) holds) then OLS is 
equivalent to fitting the model using Maximum Likelihood.
In this case a great deal of theory exists for how the OLS estimators behave statistically.
A well developed suite of diagnostic devices exists for spotting when the assumptions are violated.
Diagnostic plots and diagnostic statistics followed by omission of the offending points can be used to tackle outlier problems.
However as stressed by one of the anonymous referees ``simple strategies like
outlier detection and elimination based on diagnostic plots tend
to work reliably only in simple regression with one predictor and
a few outliers - but even then problems arise because little can
be said about the distribution of the final parameter estimators
based on such deletion procedures, so that the use of such
estimators should not be encouraged.''

OLS is also not very robust to the other
complications we might reasonably expect to crop up in astronomical data, Section~\ref{sec:features}. For example, when there is error on the predictor variable (i.e. $x^{obs}$ is a noisy
version of $x$) which is uncorrelated with the error on the response variable, then the estimate of the slope of the regression line $a$ is statistically biased towards zero (see, for example, \cite{buonaccorsi2009measurement}, Chapter 4). This
has an obvious negative effect when we are primarily interested in assessing scaling relations. As a result, various
alternatives have been developed and we describe these in the following sections.

\subsection{Weighted least squares fits} 
\label{sec:ols}

OLS minimises the sum of the squared residuals of fitting the line
$ax+ b$ to the data $y^{obs}$. If the $y$ errors are homoscedastic
(i.e. all with variance equal to $\sigma^2$),
minimising the sum of the squared residuals is equivalent
to minimising
\begin{equation}
\sum^N_{i=1} \frac{ (y^{obs}_i-(b+ a x_i))^2}{\sigma^2}   .
\label{eqn:case1}
\end{equation}
When there is intrinsic scatter and the observation noise is actually heteroscedastic, 
the logical next step in
many astronomical works is to minimise a weighted sum of residuals
\begin{equation}
\chi^2 = \sum^N_{i=1} \frac{(y_i^{obs}-(b+ a x_i))^2}{\sigma^2_{i}} 
\label{chi2}
\end{equation}
where $\sigma^2_{i} = \sigma^2_{scat} + \sigma^2_{y,i}$ represents the intrinsic scatter plus the
heteroscedastic measurement noise in the notation of Equation~(\ref{eqn:full}) and these values are assumed known.
See \cite{press2007numerical} for numerical details of minimising this penalty.

Of course, we are also likely to be facing observational errors on the $\{x_i\}$, so that in fact
we observe instead the $\{x^{obs}\}$. 
If the observation noise on both variables and the intrinsic scatter are statistically independent
and known,
one may introduce a new reweighted penalty
\begin{eqnarray}
\chi^2 &=& \sum^N_{i=1} \frac{(y^{obs}_i-(b+ a x^{obs}_i))^2}{\sigma^2_{i}+a^2 \sigma^2_{x,i}} .
\label{genchi2}
\end{eqnarray}
We stress that the standard notation $\chi^2$ does not
imply that the statistic follows the distribution of the same name.
This change in weighting arises because following Equation~(\ref{eqn:full}),
\begin{eqnarray}
y_i^{obs} & = & y_i + \epsilon_{y_i} 
\mbox{ where } \epsilon_{y_i} \sim N( 0 , \sigma^2_{y,i})  \nonumber \\
& = & (a x_i + b + \epsilon_{scat} ) + \epsilon_{y_i} 
\mbox{ where } 
\epsilon_{scat} \sim N( 0 , \sigma^2_{scat}) \nonumber  \\
& = & a (x_i^{obs} - \epsilon_{x_i} ) + b + \epsilon_{scat}  + \epsilon_{y_i} 
\mbox{ where } \epsilon_{x_i} \sim N( 0 , \sigma^2_{x,i})  \nonumber \\
\mbox{so } Var( y_i^{obs} ) & = & a^2 \sigma^2_{x,i} + \sigma^2_i \ \mbox{where as before } \sigma^2_i = \sigma^2_{scat} + \sigma^2_{y,i}.
\end{eqnarray}
The optimisation is complicated by the $a^2$ term on the
denominator which makes the problem nonlinear, \cite{press2007numerical} provides a suitable
numerical optimisation algorithm known as FitEXY. 
Minimisation of this reweighted penalty requires that the $\{\sigma^2_i\}$ and
$\{\sigma^2_{x,i}\}$ are known. 
Weighted least squares fits cannot address an unknown intrinsic scatter by including 
the parameter when minimising  over all the unknown parameters (if
$\sigma^2_{scat}$ is unknown, then doing this, the minimal $\chi^2$ obviously occurs for
$\sigma^2_{scat}\rightarrow \infty$). 
\cite{kelly2007some} points out that finding associated
statistical properties of this estimator can be achieved via bootstrapping.

\subsection{Maximum Likelihood Estimation}
\label{MLE}

Putting a greater emphasis on parametric modelling than does OLS, Maximum Likelihood Estimates (MLEs) are those parameter estimates which maximise the likelihood of the observed data.
They are much favoured by statisticians not least because of the wealth of associated theory which yields (asymptotic) sampling distributions in many cases and thus interval estimates and hypothesis tests.

In several of the scenarios considered in the previous two sections, MLE and OLS or 
weighted least squares estimates of the regression parameters $a$ and $b$
would coincide.
For example, reconsidering the case where the $\{\sigma^2_{i}\}$ are 
known, there is no
observation noise on the $\{x_i\}$, and the Normality
assumptions of Equation~(\ref{eqn:full}) hold, then the likelihood for
for $a$ and $b$ is
\begin{equation}
L( a , b ) = \prod_{i=1}^N 
\frac{1}
{\sqrt{2\pi \sigma^2_{i}}}
\exp\left ( -\frac{ (y_i^{obs} - a x_i-b)^2}{2 \sigma^2_{i} } \right ).
\label{eqn:marglike2}
\end{equation}
It is clear that the maximisers of Equation~(\ref{eqn:marglike2}) are also the
minimisers of $\chi^2$ defined in Equation~(\ref{chi2})
(\cite{cash1979parameter}).

The situation is more
complicated if the intrinsic scatter, $\sigma^2_{scat}$, has to be
determined from the data or if there is observation noise on the $\{x_i\}$;
this latter errors-in-variables regression has been considered in the statistics 
literature; for example, \cite{carroll2006measurement}, Chapter 3, and \cite{casella2002statistical}, Chapter 12, have nice discussions 
of different approaches to the problem.
In the case where 
the errors for both $x$ and $y$ are normal and homoscedastic,
the complete likelihood can be written
\begin{equation}
L( a , b , \{x_i\} , \sigma^2_x , \sigma^2_y ) = ( 2\pi\sqrt{\sigma^2_x\sigma^2_y} )^{-N}
\exp( -\frac{\sum_{i=1}^N (x_i^{obs} - x_i)^2}{2\sigma^2_x} )
\exp( -\frac{\sum_{i=1}^N (y_i^{obs} - a x_i-b)^2}{2\sigma^2_y} )
\label{eqn:biglike}
\end{equation}
where $\sigma^2_x$ represents the variance of the $x$ error
and $\sigma^2_y$ represents the sum of $\sigma^2_{scat}$ and the common variance for the
$y$ observations. They point out that by taking each $x_i = x_i^{obs}$ and letting
$\sigma^2_x \rightarrow 0$, the likelihood can be made arbitrarily large. As a result,
maximum likelihood tends only to be performed after making additional assumptions about
the variances, most commonly that $\sigma^2_x = \lambda \sigma^2_y$ where $\lambda$ is a
known positive constant (although this assumption is not without its problems,
\cite{carroll1996use}). Alternatively, if we were to assume known heteroscedastic variances
$\sigma^2_{x,i}$ and $\sigma^2_{y,i}$, the corresponding likelihood would be
\begin{equation}
L( a , b , \{x_i\} , \sigma^2_{scat} ) = \frac{(2\pi)^{-N}}{\prod_{i=1}^N \sqrt{\sigma^2_{x,i}(\sigma^2_{y,i}+\sigma^2_{scat})}} 
\exp( -\frac{\sum_{i=1}^N (x_i^{obs} - x_i)^2}{2\sigma^2_{x,i}} )
\exp( -\frac{\sum_{i=1}^N (y_i^{obs} - a x_i-b)^2}{2(\sigma^2_{y,i}+\sigma^2_{scat})} )
\label{eqn:biglike2}
\end{equation}
which could be maximised numerically over the $N+3$ unknown parameters, although again it is not clear whether
the known variances assumption is justifiable.
Equation~(\ref{eqn:biglike2}) has the unappealing property that the number of unknowns, $N+3$, increases as the number of data points, $\{x_i^{obs},y_i^{obs}\}_{i=1}^N$, increases.
In particular, this means that the usual asymptotic distributional results for MLEs do not hold.

\subsection{Robust estimation}
Parameter estimation can be affected by outliers and other violations of parametric assumptions.
Robustness to outliers and to some aspects of the Normality assumptions can be introduced 
by a switch from {\it least squares} to {\it least absolute deviations}, replacing the standard 
sum of squares by (in its simplest case)
\begin{equation}
S = \sum_{i=1}^N | y_i^{obs} - (b + a x_i)|.
\label{eqn:absval}
\end{equation}
This form is quite commonly used in astronomy, being offered in \cite{press2007numerical}; 
see also \cite{Gott} for a witty introduction.

The fitted value $\hat{b}$ is here the median of the $y_i^{obs} - (b + a x_i)$ and the estimate of the slope can be found numerically as the solution of the equation
\begin{equation}
0 = \sum_{i=1}^N x_i \  \mbox{sign}( y_i^{obs} - (\hat{b} + a x_i) ).
\end{equation}
(There may not be a unique solution to this problem.)
Least absolute deviations regression is quite robust against unusual $y_i^{obs}$ but it does not cope so well with unusual $x_i$ values (high leverage points) 
and can be quite unstable with regard to small changes in the $\{x_i\}$. 
However it is just one possible robust approach and this is an active field of research, see for example
\cite{huber2009robust}, Chapter 7.

\subsection{Bivariate Correlated Errors and Intrinsic Scatter (BCES)}

In 1996 \cite{akritas1996linear} proposed 
a widely applicable extension of OLS, known as BCES. 
It considers
regressions with heteroscedastic and even correlated errors on
both axis as well as an intrinsic scatter.
Again the assumption is made that the measurement error
variances and covariances are known, although they may vary from point to point, and importantly
they may be a function of the response and predictor variables. 
The BCES estimator extends OLS by
considering the usual estimates in terms of the moments of the bivariate distribution of $x$ and
$y$, Equation~(\ref{eqn:OLS}), and then expressing these moments in terms of the moments of the
observed $x^{obs}$ and $y^{obs}$ and the known measurement variances.
In analogy with Equation~(\ref{eqn:OLS}), the BCES estimates of slope and intercept when regressing $y$ on $x$ are
\begin{eqnarray}
\hat{a} & = & \frac{\widehat{\mathsf{Cov}}(x^{obs},y^{obs})-\frac{1}{N} \sum_{i=1}^N \sigma^2_{xy,i} }{\widehat{\mathsf{Var}}(x^{obs})-\frac{1}{N} \sum_{i=1}^N \sigma^2_{x,i}} \nonumber \\
\hat{b} & = & \bar{y}^{obs} - \hat{a} \bar{ x}^{obs}
\label{eqn:BCES}
\end{eqnarray}
where $\sigma^2_{xy,i}$ is the known covariance between $\epsilon_{x,i}$ and $\epsilon_{y,i}$ and $\sigma^2_{x,i}$ is the known variance of $\epsilon_{x,i}$ (as defined by Equation~(\ref{eqn:full})).
These estimates are referred to as the ``Moment-based corrected estimators'' in \cite{buonaccorsi2009measurement}, Chapter 4, where fuller discussion is given regarding their properties.
Asymptotically, the
slope and intercept estimators are shown by \cite{akritas1996linear} to have Gaussian sampling distributions and so confidence
intervals can be estimated.
As an alternative, BCES also assesses the fitting errors using bootstrap ideas (repeatedly resampling from the observed data to form an empirical sampling distribution of the estimators).
See the original paper for more complete details.

Equation (\ref{eqn:BCES}) provides point estimates for the regression of $y$ on $x$. 
\cite{akritas1996linear} also derive estimates for when the regression of $x$ on $y$ is preferred.
As the two resulting lines will not be equal, the BCES bisector is also defined as the bisector of the two.

BCES addresses some of the difficulties arising in scaling
relations in astrophysics, namely errors on measurements which
are potentially correlated as well as being functions of the true
$x$ and $y$. 
However, problems
related to outliers, data structure, upper-limits and selection effects are not addressed by
this methodology, and in small data sets non-Normality of the data may be an issue for exact confidence intervals. 
Code to implement these estimators is available
via
\newline
\texttt{http://www.astro.wisc.edu/$\sim$mab/archive/stats/stats.html}.

\subsection{Astronomy Survival Analysis (ASURV)} 
\label{sec:asurv}
In the wider world of statistics, there
are many application areas in which data are censored and the field of survival analysis
has developed to deal with such data (see \cite{miller1981survival} for example). As the
name perhaps suggests, survival analysis tends to deal with cases where observations can
only be made when the subject survives more or less time than a certain observation date.
As such, the analogy with flux or other attribute limited selection is quite clear.
The analogy is not perfect however  when we are thinking about upper/lower limits or
selection function, because these are stochastic (the inclusion/exclusion occurs with
some probability), while censoring is deterministic.

\cite{schmitt1985statistical} and \cite{isobe1986statistical} introduced the  astronomical
community to these ideas thereby allowing astronomers to perform regression in the
presence of upper/lower limits (following the paper \cite{feigelson1985statistical} on a
similar theme for univariate problems).
\cite{schmitt1985statistical} leads the reader through a series of statistical procedures in the
presence of censoring, of which we will concentrate here on regression with the assumption of
upper-tail censoring on both variables and a known number of potential observations. The suggested
procedure begins by finding a nonparametric estimate of the joint distribution of the two
variables. To do this, the data are first binned on each variable separately, including a bin for
the censored values. Then considering the intersection of the two sets of bins, four distinct
classes emerge: points for which both measurements are available, points for which only one
variable is available with the other censored (the two cases), and points for which both values
are censored. Based on counts of the numbers of points in each class,
\cite{schmitt1985statistical} derives Maximum Likelihood estimates of the probability of each bin
combination. This stage does not make any parametric assumptions about how the data are distributed.
Finally, given the estimated joint distribution, the moments of the distribution can be found and
used in the usual OLS formulations, Equation~(\ref{eqn:OLS}). 
See the original paper for more complete details.

\cite{isobe1986statistical} reviews the methodology of \cite{schmitt1985statistical}, adding two further approaches.
In the first, data are assumed only to be censored on the response variable, and a distributional assumption of Normality is made for the response variable then using the EM algorithm \cite{dempster1977maximum} to impute the censored values for use in a Maximum Likelihood approach.
In the second, a nonparametric approach is taken, relaxing the Normality assumption but again only assuming censoring on one variable, making it perhaps less widely applicable than the approach of \cite{schmitt1985statistical}. 
The paper compares the various approaches via simulation studies and on several astronomical applications.
A useful resource written by some of the authors of this paper, amongst others, is code to implement these estimators, available at \texttt{http://astrostatistics.psu.edu/statcodes/asurv} and described in \cite{lavalley1992asurv}.

The explicit handling of censoring greatly extends the range of regression problems which can be
tackled appropriately, as does the nonparametric approach of some of the methods. However some
limitations remain; there is not currently provision for incorporating intrinsic scatter nor
heteroscedastic error structure, nor the other common features of astronomical
data in Section~\ref{sec:features}.
In particular, ASURV only considers a specific type of upper limit censoring,
which is a sharp upper limit.
As already mentioned, the most usual case in astronomy is a soft,
probabilistic, threshold,
one in which also some objects below the threshold
may be observed while some objects above the threshold will 
be missed.

\subsection{Errors-in-variable Bayesian regression}
\label{sec:Bayes}

Bayesian regression started to receive significant attention in the astronomical community after
the publication of \cite{d2005fits}, which is a brief summary of the relevant content in
\cite{d2003bayesian}, and the errors-in-variable model of \cite{dellaportas1995bayesian}. The
important differences for the practitioner between Bayesian and non-Bayesian approaches lie in
the model specification and interpretation of results. In terms of the former, a Bayesian approach
takes the usual question ``How did the data I observed arise?'' and adds the new question
``What do I know about the parameters even before I collect any data?''. 
Effectively it asks the practitioner to quantify
their uncertainty about the parameters in the form of a prior distribution
(which may be either very precise or, more commonly, very vague). This is combined with the usual
likelihood for how the data arise via Bayes' theorem to form a posterior distribution which now
quantifies our uncertainty about the parameter having observed data. 
In a mathematical sense, the posterior distribution is simply proportional to the product of the likelihood function
and the prior distribution.
Summarising the posterior by its location (mean, median, mode, etc) 
quantifies the parameter value; its spread, e.g. the interval that includes 68\% of it, quantifies the parameter uncertainty. 
On the practical side, there are now a number of Bayesian packages which
allow the user to specify relatively general models without needing to compute the posterior in closed
form (i.e. as a single mathematical expression whose derivation is often difficult).
Amongst them, perhaps the most notably are BUGS
\cite{lunn2009bugs} and JAGS \cite{plummer2010jags}.

Despite the philosophical differences, since a Bayesian approach and fitting by Maximum Likelihood share a common model for how the data arise, it is often the case that, numerically, the two sets of resulting point estimates can be quite close, especially if there is a lot of informative data or if the prior information is weak.
This is entirely positive.
Where a Bayesian analysis may really help is when perhaps we have strong prior information or wish to explore the uncertainty associated with relaxing some condition required for Maximum Likelihood estimation (for example, the constant ratio of variances assumption).  
It may also permit some dimension reduction if the latent variables $\{x_i\}$ are not of interest by integrating them out of the posterior distribution analytically, as is done by \cite{d2005fits}.
For example, considering Equation~(\ref{eqn:biglike2}), the posterior for  just $a,b,\sigma^2_{scat}$ can be expressed 
\begin{eqnarray}
p( a,b,\sigma^2_{scat} | \{x_i^{obs},y_i^{obs}\}) & = & \int_{\{x_i\}}
p( a,b,\{x_i\},\sigma^2_{scat} | \{x_i^{obs},y_i^{obs}\})\ d\{x_i\}  \nonumber\\
& \propto & \int_{\{x_i\}} L(a,b,\{x_i\},\sigma^2_{scat})
p( \{x_i\}  ,a, b)\  d\{x_i\}
\label{eqn:marglike}
\end{eqnarray}
where $p( \{x_i\}  ,a, b)$ is the joint prior and the integration is over the $N$ dimensional latent variables.
Obviously this dimension reduction approach will only be feasible in certain 
analytically tractable cases.  When analytic dimension reduction is
not an option, numerical marginalisation is required, either directly by numerical integration
or indirectly by Monte Carlo methods.

To illustrate the Bayesian approach, a version of the errors-in-variable regression model assuming known $\{\sigma^2_{x,i}\}$ and $\{\sigma^2_{y,i}\}$ takes the usual likelihood equations
\begin{eqnarray}
x^{obs}_{i}  \sim & N(x_i, \sigma^2_{x,i}) & \mbox{Gaussian errors on } x^{obs} \nonumber \\
y^{obs}_{i}  \sim & N(y_i, \sigma^2_{y,i}) & \mbox{Gaussian errors on } y^{obs}\nonumber \\ 
\mbox{where }y_{i}  \sim & N( a x_i + b , \sigma_{scat}^2) & \mbox{Gaussian intrinsic scatter around a straight line for } y
\label{eqn:ex_lik}
\end{eqnarray}
and adds the equations specifying the additional question ``What do I know about the parameters even before I collect 
any data?'', for example, we could specify that independently
\begin{eqnarray} 
x_i  \sim & U(-10^4, 10^4) & \mbox{Uniform population structure for }x_i,\ i=1,\ldots,N \nonumber \\
\sigma^2_{scat} \sim& InvGam (0.01,0.01) &  \mbox{Inverse Gamma prior on scatter}\nonumber  \\
a \sim& t_1 & \mbox{Student's t on slope, equivalent to a uniform on angle} \nonumber \\ 
b \sim& N(0,10^4) & \mbox{Gaussian prior on intercept} .
\label{eqn:ex_p1}
\end{eqnarray}
(The choice of an Inverse Gamma prior for the variance term is quite common but is not without 
criticism, \cite{gelman2006prior}. The numerical values in the uniform prior are meant to indicate realistic limits for a particular application rather than being general purpose.)
Precise information could be introduced for particular applications.
For example we may know from previous
analyses that the slope is $\pi\pm0.1$, and
in such a case we could use $a \sim N(\pi, 0.1^2)$;
the prior is one way to take advantage of the work
of previous generations of researchers.  
Code to implement the error-in-variable Bayesian regression is given in the Appendix.
Other
common features of astronomical data can be addressed, as we detail
in Section~\ref{sec:bayes.full}.

Turning to one of the other uses of regression mentioned in Section~\ref{sec:regress}, prediction,
before data $z$ are collected (or even considered), the distribution of 
predicted values $\widetilde{z}$ can be expressed as
\begin{equation} 
p( \widetilde{z} ) = \int p( \widetilde{z}, \theta ) d \theta = \int  p( \widetilde{z} 
| \theta ) p(\theta) d \theta .
\label{eq.pred}
\end{equation} 
These two equalities result from the application of probability
definitions, the first is simply that a marginal distribution results
from integrating over a joint distribution,
the second is Bayes' rule. 
If some data $z$ have been already collected for similar objects, we can
use these data to improve our prediction for $\widetilde{z}$.
For example, if mass and richness in clusters are highly correlated, 
one may better predict
the cluster mass knowing its richness than without such 
information simply because
mass shows a lower scatter at a given richness than when
clusters of all richnesses are considered (except if the relationship
has slope exactly equal to $\tan k \pi/2$, with $k=0,1,2,3$). 
In making explicit the presence of such data, $z$,
we rewrite Equation~(\ref{eq.pred}) conditioning on $z$
\begin{equation} 
p( \widetilde{z} | z ) = \int p( \widetilde{z} | z, \theta ) p(\theta|z) d \theta  .
\label{eqn:Bpredict}
\end{equation} 
The conditioning on $z$ in the first term in the integral simplifies 
because $z$ and $\widetilde{z}$ are considered conditionally independent given
$\theta$, so that this term becomes $p( \widetilde{z} | \theta )$. The left
hand side of the equation is called the posterior predictive distribution 
for a
new unobserved $\widetilde{z}$ given observed data $z$ and model parameters
$\theta$.  Its width is a measure of the uncertainty of the predicted value
$\widetilde{z}$, a narrower distribution indicating a more precise prediction. 
Examples are given in Section~\ref{sec:perform}.

\section{Performance comparisons}
\label{sec:perform}

In this section we consider existing and new comparisons of some of the different regression techniques.
Most comparisons tend to use
relatively simple simulated data sets because many of the regression techniques
do not address more than a few of
the features of astronomical data listed in Section~\ref{sec:features}.
We split our review into performance in estimating the regression and performance in prediction.
We have not included ASURV in the comparisons because it addresses 
only censoring of all the possible data complications and this is
relatively uncommon in astronomy, which deal instead mostly with a soft,
probabilistic, threshold.

\subsection{Estimating the regression parameters}

\cite{kelly2007some} compares
performance in recovering the regression parameters for 
OLS, BCES and FitEXY and a likelihood model he calls the Gaussian structural model proposed in that paper.
For OLS and FitEXY poor performances might be anticipated
because they are applied outside their range of validity since
the simulated data has heterogeneous errors on both $x$ and $y$ (not
addressed by OLS) and
a unknown intrinsic scatter (not addressed by FitEXY). 
In finding the Maximum Likelihood Estimates based on the Gaussian structural model he compares asymptotic theory to derive
uncertainty estimates with a semi-Bayesian approach drawing samples from the associated posterior when using very weak priors.
\cite{kelly2007some} found that OLS returns
slopes which are statistically biased towards zero, in agreement with \cite{akritas1996linear}, 
while BCES and FitEXY return
intrinsic dispersions which are systematically wrong. This is unsurprising
for FitEXY and OLS since they do not address the simulated case (we
emphasise that FitEXY allows an intrinsic scatter only
if known).
The Gaussian structural model in its semi-Bayesian approach (ignoring the prior 
when computing the point estimate of the parameters, but using 
it to compute their uncertainties) outperforms
the other methods, including BCES.

\begin{figure}
\centerline{
\subfloat[]{\includegraphics[clip, width=5.5truecm]{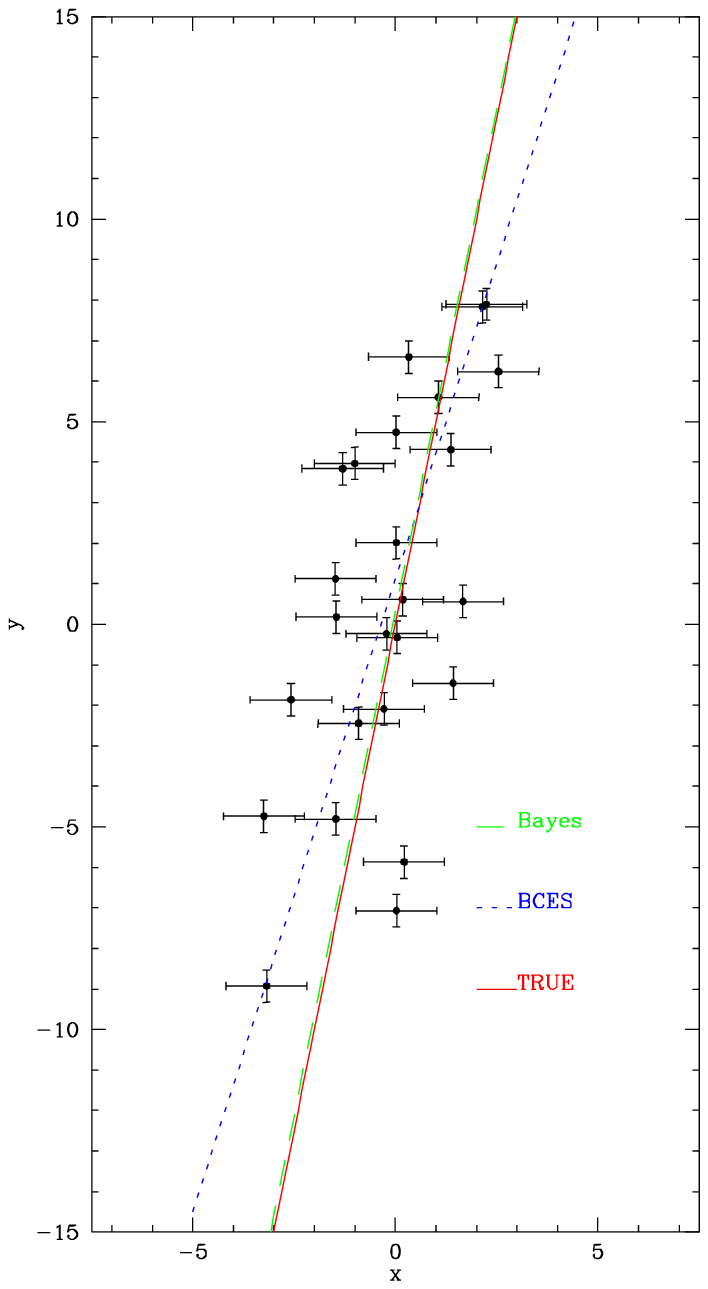}}
\subfloat[]{\includegraphics[clip, width=5.5truecm]{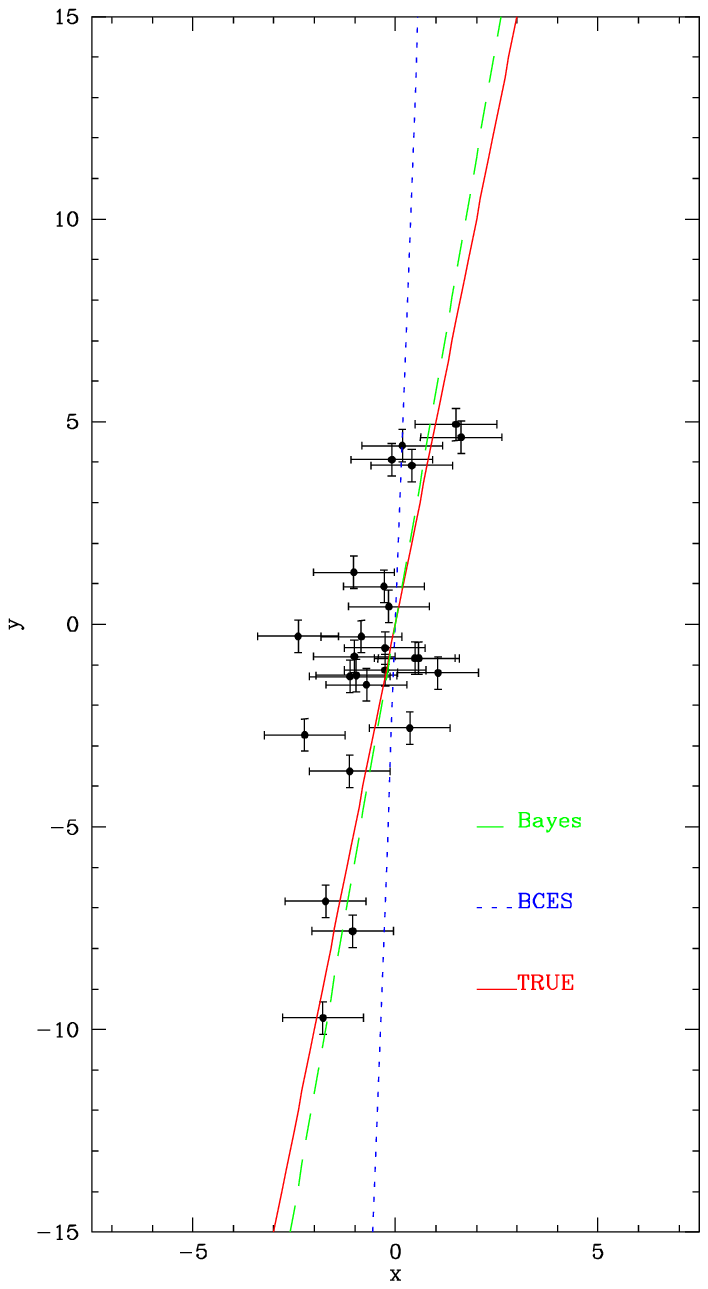}}
\subfloat[]{\includegraphics[clip, width=5.5truecm]{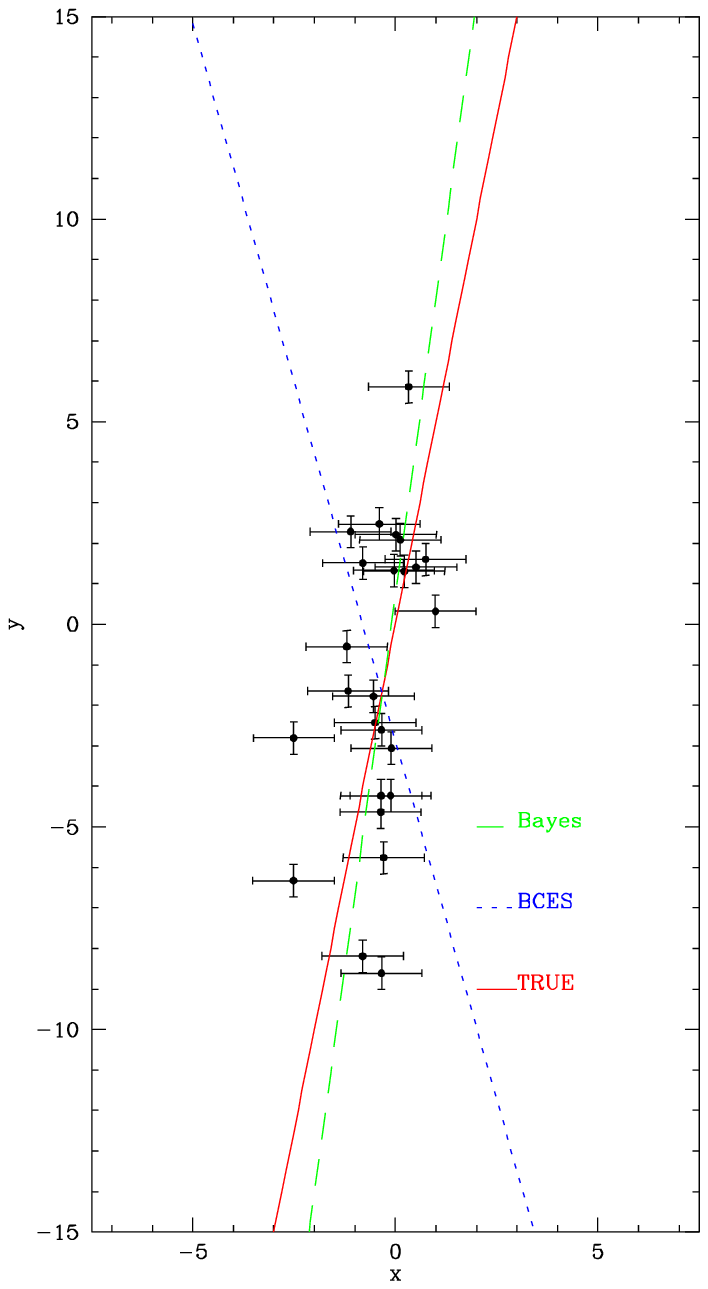}}
}
\caption{Three simulated data sets each of 25 points, showing
true trends from which the data are generated in red, the
trend recovered by BCES in blue, and the mean Bayesian estimate in
green. 
In 1000 simulations, BCES results performed worse than those shown in
the central and right panels about 10\% of the times.
The simulations are meant to mimic the $L_X-\sigma$
scaling of galaxy clusters (from \cite{andreon2010bayesian}).
}
\label{showregr}
\end{figure}

\begin{figure}
\centerline{
\includegraphics[clip, width=12truecm]{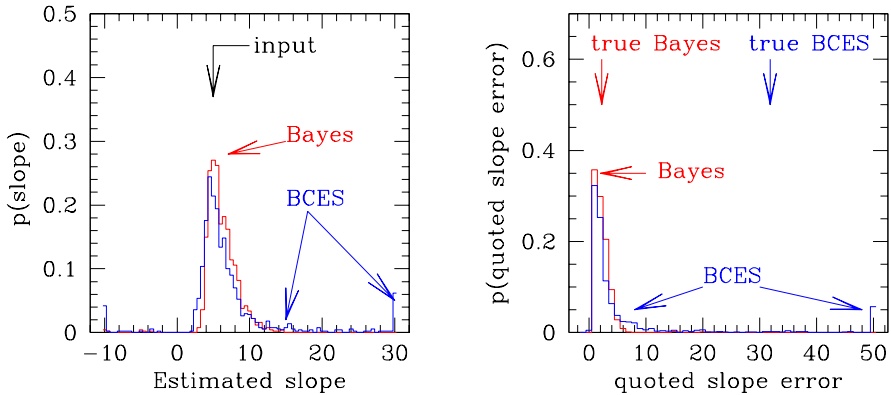}
}
\caption{
Comparison between BCES (blue) and the Bayesian
approach (red) for a linear regression problem (from \cite{andreon2010bayesian}) using 
1000 simulations of a sample of 25 objects.
{\it Left panel:} Histogram of point estimates of 
the slope parameter, note that the extreme bins represent $\le -10$ and $\ge30$. In approximately 12\% of the simulations, BCES  
returns significantly wrong estimates,
the fit associated with one of these is plotted in the right hand panel of Figure~\ref{showregr}.
{\it Right panel:} Histogram of estimated variability of slope estimates; for BCES this is the returned standard error, for the Bayesian approach it is the returned standard deviation of the posterior mean for the slope.
Note that estimates larger
than $50$ are binned for display purposes at $50$.
The two arrows denote the standard deviations calculated from the 1000 simulations of the observed differences between the true and the estimated slopes.
}
\label{regrperformance}
\end{figure}

\cite{andreon2010bayesian} compares two methods only, BCES and the 
errors-in-variable 
Bayesian model \cite{dellaportas1995bayesian} of Section~\ref{sec:Bayes}.
In this case, the comparison uses models within the 
range of validity, generating 1000 samples of 25 data points drawn from a linear
regression with
slope $a=5$, an intrinsic scatter $\sigma^2_{scat}=1$, and
homogeneous Gaussian errors with $\sigma^2_x=1$ and $\sigma^2_y=0.4^2$.
These parameters
are chosen to approximate X-ray luminosity vs velocity dispersion of
clusters of galaxies e.g. \cite{Aetal08}. 
Three examples are shown in Figure~\ref{showregr}. \cite{andreon2010bayesian} finds that
BCES sometimes estimates the slope parameter very poorly (see the right hand panel in
Figure~\ref{showregr} for an example and the left hand panel in
Figure~\ref{regrperformance} for an ensemble view).
When BCES does estimate the slope poorly, it does also
return a much larger estimated measure of error, right hand panel in
Figure~\ref{regrperformance}.
On average over the 1000 simulations, the Bayesian approach 
has a smaller measure of error than BCES.

Computationally, the costs of the different techniques vary considerably, OLS being the cheapest and the Bayesian errors-in-variables the most expensive as it requires Markov chain Monte Carlo methods (MCMC).
In this later case, depending on the size of the data set, the cost may be up to a few minutes 
or hours depending on the complexity of the model rather than a few seconds.

\subsection{Prediction}
\label{sec:pred}

Turning to a comparison of prediction accuracy,
we consider a simulation study involving a simpler scenario 
than that in the previous section.
We generate 100 values of $x_i$ from a non-central scaled Student-t distribution with 10 degrees of freedom,
location $0.1$, and scale $1.1$.
These $x_i$ are perturbed by homoscedastic Gaussian noise with $\sigma_x^2=1$ 
to give 100 $x_i^{obs}$.
The regression line takes the form $y=x$ with no intrinsic scatter, and
these $y_i$ are also perturbed by homoscedastic Gaussian noise with $\sigma_y^2=1$ 
giving 100 $y_i^{obs}$. 
The resulting $\{x_i^{obs},y_i^{obs}\}$ pairs are used as the data
for
the errors-in-variable Bayesian model, OLS, weighted least squares, BCES and MLE (in all cases
the variances of the noise are assumed known). 
The likelihood for the unknowns $a,b,\{x_i\}$ based on the observations
$\{x_i^{obs},y_i^{obs}\}$ can be written
\begin{equation}
L( a , b , \{x_i\} ) = \frac{(2\pi)^{-N}}{\prod_{i=1}^N \sqrt{\sigma^2_{x} \ \sigma^2_{y}}}
\exp( -\frac{\sum_{i=1}^N (x_i^{obs} - x_i)^2}{2\sigma^2_{x}} )
\exp( -\frac{\sum_{i=1}^N (y_i^{obs} - a x_i-b)^2}{2\sigma^2_{y}} )
\label{eqn:biglike3}
\end{equation}
Noting that for MLE for this experiment the likelihood has 102 unknown 
parameters. We follow \cite{d2005fits} in using the likelihood
\begin{equation}
L( a , b ) = \frac{\sqrt{2\pi}^{-N}}{\prod_{i=1}^N \sqrt{\sigma^2_{y}+a^2\sigma^2_{x}}}
\exp( -\frac{\sum_{i=1}^N (y_i^{obs} - a x_i^{obs}-b)^2}{2(\sigma^2_{y}+a^2\sigma^2_{x})} )
\label{eqn:dagolike}
\end{equation}
which corresponds to integrating out the $\{x_i\}$. 
Notice the correspondence between the exponential term in this equation and 
Equation~(\ref{genchi2}), although they differ in that Equation~(\ref{eqn:dagolike}) also has an $a$ dependence in the leading non-exponential term.
Therefore, when there are errors on the predictor quantity, 
weighted least squares is not the MLE estimate.

For the Bayesian errors-in-variable model we retain the $\{x_i\}$ in the model and so will need to assume a prior population
model for them.
Retaining the $\{x_i\}$ in the model does increase the computational cost although as a byproduct it does also generate estimates of the $\{x_i\}$ which might be useful in some applications.
In some astronomical problems, the $x$ data structure is fairly well known
(e.g. from previous experiments or from theory). 
If this were the case here, the prior would be
\begin{equation}
x_i \sim t_{10}(0.1, 1.1^2)
,\  i=1,\ldots,100
\end{equation}
since this is actually the true population model.
Of course it may well be the case that we do not have such a high level of prior knowledge 
and so we also consider an alternative prior
\begin{equation}
x_i \sim N(0, 1^2) ,\  i=1,\ldots,100
\end{equation}
which has a difference in location and scale as well as a lighter
tail behaviour as an example of possible mismatch between
the true data structure and what we know about it. 
This prior is still relatively informative; it might perhaps be arrived at in a not strictly Bayesian sense by looking at a histogram of the 100
$x_i^{obs}$ values.
Finally, we also consider a less informative prior, with parameters
to be determined at the same time as the other regression parameters;
we have adopted a Normal prior with parameters treated as additional unknowns
\begin{equation}
x_i \sim N(\mu_{prior}, \sigma^2_{prior}) ,\  i=1,\ldots,100
\end{equation}
with weak hyperpriors on the parameters
\begin{eqnarray}
\mu_{prior} &\sim& N(0,10^4) \\
1/\sigma^2_{prior} &\sim& U(0,10) .
\end{eqnarray}

Using the observed data, the Bayesian model makes predictions of $y$, $\widetilde{y}$,
for new $x^{obs}$ using Equation~(\ref{eqn:Bpredict}). 
The non-Bayesian
techniques have point estimates of the slope $\widehat{a}$ and intercept
$\widehat{b}$ using the original 100 points, and these are used to
give predictions $\widetilde{y}=\widehat{a} x^{obs} + \widehat{b}$ for
new $x^{obs}$. To assess performance, $10000$ new $x_i^{obs}$ values are
simulated from the model and for each we calculate the residual between
the predicted $\widetilde{y_i}$ and the actual $y_i$.
Figure~\ref{regr.competitors.dt} plots these residuals against
the value of $x^{obs}$ for the various different fitting algorithms
(binning the results in small $x^{obs}$ intervals to aid visualisation).
Notice that we plot the residuals against $x^{obs}$ rather than against $x$;
although this induces a correlation with the residuals for some methods, it is actually the performance for an observed $x^{obs}$ rather than an unobserved $x$
which may be of interest to practitioners.
This also emphasises the point made by \cite{carroll2006measurement}, Chapter 2, that 
homoscedastic measurement error is less of a problem for prediction than it is for 
parameter estimation because one is predicting on the basis of $x^{obs}$ and not $x$ 
and thus methods which ignore the distinction are not penalised to the same extent.
In this simulation, that makes OLS a competitive standard.

Running through the results as laid out in Figure~\ref{regr.competitors.dt}:

\begin{figure}
\centerline{
\includegraphics[clip, width=10truecm]{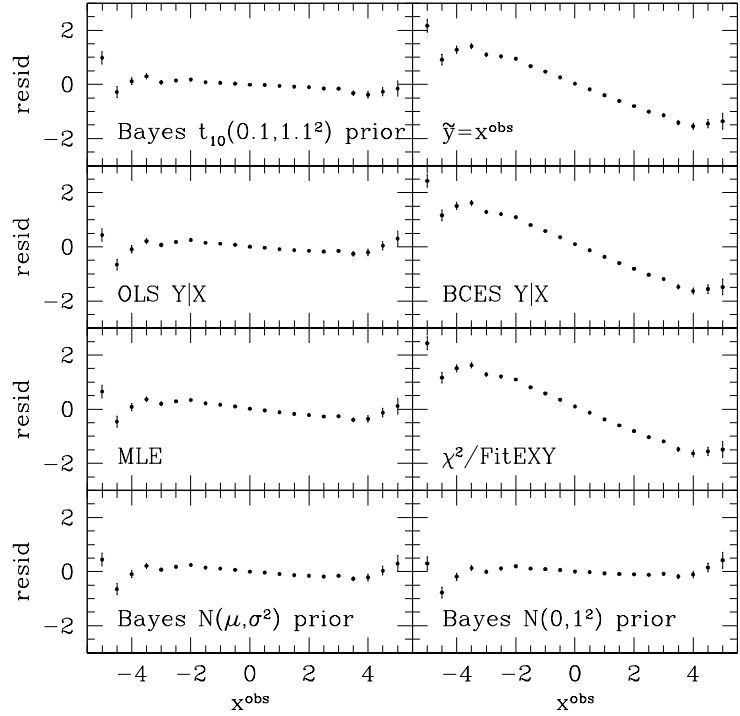}
}
\caption{
Performance of the various approaches in predicting
the true value $y$ given $x^{obs}$. Each point show the average residual
per small bin of $x^{obs}$. 
The error bars indicate the standard error of the mean residuals. 
\label{regr.competitors.dt}}
\end{figure}

\begin{description}

\item[Top row, left]
Bayes regression with a $t_{10}(0.1,1.1^2)$ prior
performs very well with small residuals across the entire range of
$x^{obs}$ as is to be expected given that the correct model is being fitted.
The point estimates of the parameters themselves are 1.09 and 0.00 for the slope and intercept respectively.

\item[Top row, right]
Suppose that one naively attempts
to predict $\widetilde{y}$ by $\widetilde{y}=x^{obs}$, in effect making use of the knowledge that $y=x$ but ignoring the noise structure.
The residual in this case is the perturbation $x-x^{obs}$, i.e. a $N(0,1)$ value in this example, however it is as a result negatively correlated with $x^{obs}$ so that
the largest residuals occur at the extremes of $x^{obs}$ 

\item[Second row, left]
OLS generates better predictions than the naive estimate.
It achieves this better performance
by underestimating the true slope, returning an estimated
slope of $0.65$ in place of the input $1$ and an intercept of 0.02 in place of 0, in agreement with \cite{akritas1996linear} and
Section~\ref{sec:regress}.
Feeling that this fitted line is not steep enough to capture
the data trend (see discussion on Figure~\ref{regr.dirinv} for a related case), 
some astronomical papers average this slope with
the one derived by swapping $x^{obs}$ with $y^{obs}$. 
However while this may be more appropriate for parameter estimation, it is suboptimal for prediction.
In this case this gives a
new fitted line with a slope near to one, as in our $\tilde{y}=x^{obs}$ case,
and consequently the performance of this averaged OLS method is close to
the naive estimator case.

\item[Second row, right] 
Predicted $\tilde{y}$ values computed using the BCES method 
are similar to those using the naive $\tilde{y}=x^{obs}$ method.
Comparing Equation~(\ref{eqn:BCES}) when $\sigma^2_{xy,i}=0$ with 
Equation~(\ref{eqn:OLS}) it is apparent that in improving the estimates of the
slope and intercept, the prediction performance is degraded.

\item[Third row, left] MLE using Equation~(\ref{eqn:dagolike}) returns
reasonable predictions.

\item[Third row, right] $\chi^2$/FitEXY, Equation~(\ref{genchi2}) performs less well than MLE 
because Equation~(\ref{eqn:dagolike}) has a leading term missing in Equation~(\ref{genchi2}).

\item[Bottom row, left and right]
Bayesian errors-in-variables regression with either a $N(\mu,\sigma)$ prior for the $\{x_i\}$ or a 
$N(0,1)$ prior performs almost as well as the Bayesian model using the true distribution as a prior, showing
that an accurate knowledge of the prior is not necessary to achieve 
good performance in this example. 
\end{description}

In addition to Figure~\ref{regr.competitors.dt}, Table~\ref{tab1} lists residuals at a few points ($|x^{obs}|=1,3,5$) for all our
comparison performances.
To summarise for this example the methods based on sound statistical models
applied appropriately (MLE, Bayesian regression) outperform BCES and weighted least squares.
OLS also does well here paradoxically because of its sheer simplicity.
Of these methods, the Bayesian regression also generates estimates for the $\{x_i\}$, if these are required, as well as reliable estimates of the regression parameters (unlike OLS).
If we were talking about predicting observationally expensive masses 
($\widetilde{y}$) from observationally parsimonious mass
proxies ($x^{obs}$), such as richness or X-ray luminosity,
non-Bayesian fitting models are either inaccurate in
predicting the trend between mass and proxy (i.e. return an incorrect slope
of the $y$ vs $x$ relation) or perform very badly in predicting
masses from observed values of mass proxies. The Bayesian fit
estimates the slope accurately and provides well predicted masses, and so
may achieve better results for the same costly telescope time, albeit at greater computational cost.

\begin{table}
\caption{Mean residuals at various values of $x^{obs}$}
{
\footnotesize
\begin{tabular}{l r r r r r r}
\hline
		& $x^{obs}=-5~$ & $x^{obs}=-3~$ & $x^{obs}=-1~$ & $x^{obs}=1~~~$ & $x^{obs}=3~~~$ &  $x^{obs}=5~~~$ \\
\hline
Bayes $t_{10}(0.1,1.1^2)$ prior 	&$0.98$ &$0.06$ & $0.04$ & $-0.07$ & $-0.17$ & $-0.15$ \\
$x^{obs}$		&$2.16$ &$1.10$ & $0.47$ & $-0.40$ & $-1.14$ & $-1.36$ \\ 
OLS			&$0.44$ &$0.07$ & $0.11$ & $-0.09$ & $-0.15$ & $0.30$ \\ 
BCES			&$2.43$ &$1.29$ & $0.58$ & $-0.37$ & $-1.19$ & $-1.49$ \\ 
MLE 			&$0.65$ &$0.20$ & $0.17$ & $-0.12$ & $-0.26$ & $0.11$ \\ 
$\chi^2$/FitEXY 	&$2.43$ &$1.29$ & $0.58$ & $-0.37$ & $-1.19$ & $-1.49$ \\ 
Bayes $N(\mu,\sigma^2)$ prior &$0.45$ &$0.07$ & $0.11$ & $-0.09$ & $-0.16$ & $0.29$ \\ 
Bayes $N(0,1^2)$ prior	&$0.30$&$-0.01$ & $0.09$ & $-0.06$ & $-0.08$ & $0.42$ \\ 
\hline
\end{tabular}
\hfill \break
Any uncertainty associated with the predictions for a fixed value of $x^{obs}$ is associated purely with the
uncertainty of $x$ corresponding to this $x$ (and thus the corresponding $y$). This is not an issue for
between-method comparison and so is omitted.
} 
\label{tab1}
\end{table}    

\section{Including more features of astronomical data in Bayesian regression}
\label{sec:bayes.full}

Section~\ref{sec:features} listed many common features of astronomical data. In
this section we consider how we might expand the Bayesian errors-in-variable model
to include some of these features and others. In some cases, it is a question of
changing the likelihood of how the data arise. In others, it is the prior which
must change. Notice that in the former case, a non-Bayesian approach might also
work with this same modified likelihood, usually at the cost of a more complex
fitting by Maximum Likelihood.

\begin{itemize}

\item
Heteroscedastic errors on both $x$ and $y$ are already accounted for in the errors-in-variable model.
However so far they have been assumed known.
Very often, uncertainties are not perfectly known because of the complexity of
determining them (based on properties of the mechanism by which the data are obtained, rather than by a statistical repeat sampling approach).
\cite{andreon2010scaling} extends the simple example model to allow for this uncertainty.
Denoting the value for the variance suggested by the physics by $s^2_{y,i}$, a prior is constructed which reflects both the positivity of $\sigma^2_{y,i}$ and the accuracy of $s^2_{y,i}$:
\begin{equation}
{\sigma^2_{y,i}} \sim \frac{s^2_{y,i}}{\nu}  \chi^2_\nu .
\label{eqn:ex_p2}
\end{equation}
The degrees of freedom of the distribution, $\nu$, control
the spread of the distribution, with large $\nu$ meaning that
quoted variances will be close to the true variances.
\cite{andreon2010scaling} uses $\nu=6$ to quantify
with 95\% confidence that quoted standard deviations are correct
up to a factor of 2 (i.e. 
$\frac{1}{2}<\frac{{\sigma_{y,i}}}{s_{y,i}}<2$).
Notice that it is not possible to put a flat prior on these terms since there is a lack of identifiability associated with partitioning the variability of the $y_i^{obs}$ between $\sigma^2_{y,i}$ and $\sigma^2_{scat}$.

\item
Intrinsic scatter. This is taken to be Normal in the standard errors-in-variable model,
but it could readily be replaced with something more suitable
for the particular application, for example with a Cauchy distribution,  
$y_{i}  \sim Cauchy( a x_i + b , \sigma_{scat}^2)$ if the scatter between $x$ and $y$ is thought to be exceptionally heavy tailed (the Cauchy distribution is equivalent to the $t$ on 1 degree of freedom). 
An application is provided by the computation of the color offset of stellar 
locii (\cite{AHC11}): allowing heavier-than-Gaussian tails one may account 
for contamination by QSO and objects with corrupted 
photometry.

\item 
Non-ignorable data collection and selection effects.
In this case, the problem is the same for Bayesian and non-Bayesian methods since it is the
likelihood of the observed data which is affected.  
\cite{gelman2004bayesian} nicely describes a modified likelihood taking into account selection effects.
Suppose we introduce a new binary variable $D_i$ for
observation $i$ where $D_i = 1$ if $i$ is observed and zero otherwise and $D_i$ depends only on the value of the
underlying $y_i$.
The likelihood of observing a value $y_i^{obs}$, denoting other parameters of the model collectively by $\theta$, can then
be expressed
\begin{eqnarray}
f( y_i ^ {obs} | D_i=1 , y_i , \theta ) & = & 
\frac{ f(D_i = 1 | y_i ) f( y_i ^ {obs} | y_i , \theta )}{ \int_{-\infty}^{\infty} f(D_i = 1 | z ) f( z | y_i , \theta )  dz } 
\label{eq:detect}
\end{eqnarray}
where $f( y_i ^ {obs} | y_i , \theta )$ is the usual likelihood when there are no selection effects (for example,
$N(y_i, \sigma^2_{y,i})$ in the usual errors-in-variable model) and
$f(D_i=1|y_i)$ is the sample selection function (strictly
related to the sky coverage). 
The integral in the denominator of Equation~(\ref{eq:detect}) gives the probability of
detection given the values $y_i$ and $\theta$. 
Working with this likelihood complicates both Bayesian and non-Bayesian
methods.

An application is provided by the $L_X-T$ relation of Figure~\ref{fig.LxT}, where the simulated data are generated from
the blue line, then selected according to the XMM-LSS selection function \cite{pacaud2007xmm}.
In this example, the probability that an object
enters the sample, $f(D_i=1|y_i)$, changes smoothly from zero to one depending on luminosity 
and temperature  
as it does with real data; it does not go abruptly from zero to one
at a given threshold as assumed in survival analysis.
Using the modified likelihood, the input
regression can be estimated (see \cite{AandH11} for details). 

\item
Data structure and non-uniform population. This problem is readily addressed in the Bayesian setting by choosing the most appropriate prior for the population. 
For example when dealing with the source of the Malmquist bias in faint object measurements, it might be appropriate to use
a common power law, as done in the case illustrated in Fig. \ref{fig.malmquist} (\cite{andreon2010bayesian}, see also
\cite{AB12}).

\item
Non-Gaussian data and upper limits.
Non-Gaussian data requires a change of likelihood model for $x^{obs}$, $y^{obs}$, or perhaps both.
As an example, consider the application
presented in \cite{andreon2010scaling}, illustrated in Figure~\ref{fig.scatter}(a). 
Here the observations on the $x$ axis are the logs of count data where it might be expected that count data are Poisson distributed:
If we denote the counts by $\{n^{obs}_i\}$, then a slight simplification of
the model used in \cite{andreon2010scaling}, 
would be 
\begin{eqnarray}
n^{obs}_{i}  \sim& \mathcal{P}(x_i) & \mbox{Poisson error on observation with rate } x_i \nonumber \\
x_{i}  \sim& U(0, 10^4) & \mbox{A weak prior on the underlying positive rate} \nonumber \\
y_i \sim & N( a \log(x_i) + b , \sigma^2_{scat} ) & \mbox{Gaussian regression prior on the response } y_i.
\label{eqn:ex_lik2}
\end{eqnarray}
We emphasize that
$n_i^{obs}$ is a discrete count while $x_i$ is the underlying
continuous rate.

Upper limits are automatically accounted for by the above change. Suppose, for example,
we are dealing with photons and we detected zero photons, 
$n^{obs}_{i}=0$, from a source $i$ of intensity $x_i$. This value gives an upper limit
to the source intensity (2.3 at 90\% confidence).

\item Circularity arguments. 
Very often in order to take a measurement aimed at constraining a parameter,
we need to known the parameter before performing the measurement, 
leading to a circular reasoning.
For example, if one wants to infer cosmological parameters from
clusters of galaxies, we need scaling relations (to relate the observable
quantities to mass) and 
measurements performed in standard apertures (say $r_{500}$).
However, the angular size of the latter depends on 
the cosmological parameters one is aiming to
determine.
If instead supernovae are used for the same goal
(and cosmological constraints come from the scaling between their
distance modulus and redshift, see Figure~\ref{fig.SN1A}), then one
needs to know how their absolute luminosity should be corrected (peak
luminosity is brighter for slow fading supernovae, \cite{Phillips93}),
a correction which 
in turn requires knowledge of cosmological parameters, which is
exactly what one wishes to infer.
The Bayesian solution is to model 
the dependence of the measurements on the cosmological parameters and fit
them all at once, see 
\cite{March11}, \cite{Diaferio11} and \cite{andreon2012bayesian}. 

\begin{figure}
\centerline{
\includegraphics[clip, width=8truecm]{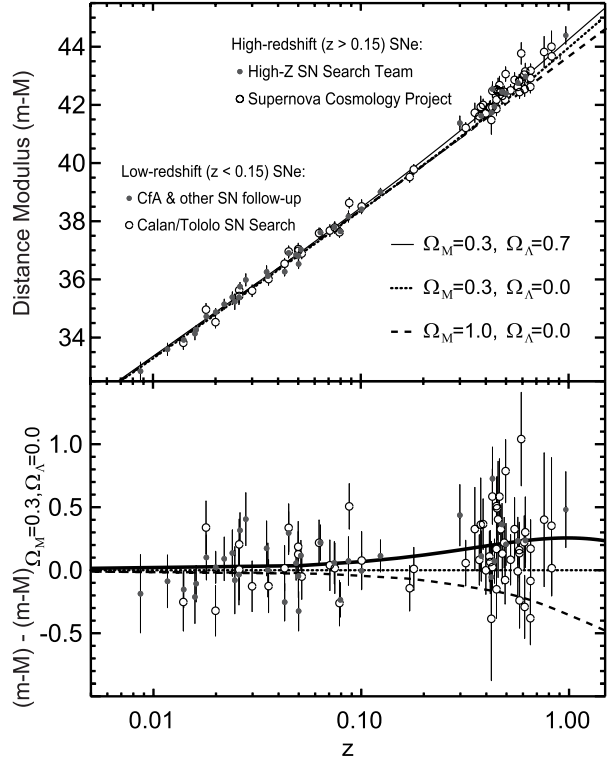}
}
\caption{Hubble diagram for supernovae. The object flux weakens 
(the distance modulus, $m-M$, becomes greater) in a way that depends on the
geometry of the Universe (from \cite{perlmutter2003measuring})
and leads to the recent Nobel prize winning discovery of  
the Dark Energy. With kind permission from Springer.
}
\label{fig.SN1A}
\end{figure}

\end{itemize}

Other complexities, such as non-linear
regressions, extra-Poisson fluctuations and contamination have been
already successfully dealt with in astronomical works using
Bayesian methods (\cite{andreon2010scaling}, \cite{AandM11}, \cite{A06}, \cite{andreon2006new})

We note that on the computational side, accounting for these and other
complications is relatively pain free in terms of statistical effort 
(if not in terms of computing time) since
converting the written symbolic equations into 
code and then implementing it is undertaken by Bayesian tools such as JAGS. 
Readers may refer to the Appendix and \cite{andreon2010scaling}, \cite{AandH11}, \cite{AandM11},  \cite{AB12},
\cite{andreon2012bayesian},  and \cite{A12metal} for examples.

\section{Conclusions}
\label{sec:conclude}

In the world of astrophysics, regression
is both important and complicated by the nature of the types of data
encountered. All approaches to regression can be seen as a way to compress the
information contained in the data into just a few numbers such as the
regression coefficients, predicted values or a measure of the support
given by the data to a model.
In this article we have attempted to review some of the
modelling problems which arise and to summarise some of the techniques
which have arisen to tackle them. We have also perhaps revealed a
personal bias towards Bayesian approaches.

Astronomy is facing an exponential increase in the number and complexity of analyses.
We are rapidly exhausting analyses not requiring complex (regression) models. 
The ease with which the Bayesian approach addresses the 
awkward features of the astronomical data makes it more and
more appealing.

\section*{Acknowledgements}
We are grateful to the anonymous referees and the Associate 
Editor for their helpful comments on an earlier version of this paper.

\appendix
\section{Appendix}
We stress that once the regression model is stated
in mathematical terms such as in Equations~(\ref{eqn:ex_lik})
and (\ref{eqn:ex_p1}), Bayesian tools such as BUGS \cite{lunn2009bugs} or JAGS \cite{plummer2010jags}
convert these equations into code and perform
the necessary stochastic computations. 
Distribution of predicted values are also a standard output of these
tools.
Equations listed in the text
find an almost literal translation in JAGS:
Poisson, Normal, Uniform and Student t distributions become
{\texttt{dpois, dnorm, dunif, dt}}, respectively. 
For some distributions, it is necessary to express them as particular cases of other distributions, for example the $\chi^2$
is a particular form of the Gamma distribution.
JAGS, following BUGS uses 
precisions, $prec = 1/\sigma^2$, in place of variances $\sigma^2$. 
Furthermore, they use natural logarithms rather than decimal ones.
The $<-$ symbol reads ``takes the value of", the $\sim$ symbol reads ``is distributed as''.
$x^y$ is coded as $pow(x,y)$ in JAGS.

\subsection{Code for the Bayesian error-in-variable regression, Section~\ref{sec:Bayes}}

{\footnotesize
\begin{verbatim}
model 
{
 for (i in 1:length(obsx))
 {
  x[i] ~ dunif(-1.0E+4,1.0E+4)  
  obsx[i] ~ dnorm(x[i],prec.x[i])
  y[i] ~ dnorm(b+a*x[i], prec.scat)
  obsy[i] ~ dnorm(y[i],prec.y[i])
 } 
 prec.scat ~ dgamma(1.0E-2,1.0E-2)
 b ~ dnorm(0.0,1.0E-4)
 a ~ dt(0,1,1)
}
\end{verbatim}
}

\subsection{Code for generating simulated data in Section~\ref{sec:pred}}

{\footnotesize
\begin{verbatim}
model 
{
 x ~ dt(0.1,pow(1.1,-2),10)
 y <- 1.0 * x + 0.0
 obsx ~ dnorm(x,1)
 obsy ~ dnorm(y,1)
}
\end{verbatim}
}

\subsection{Code for Bayesian regression model in Section~\ref{sec:pred}}

{\footnotesize
\begin{verbatim}
model 
{
 alpha ~ dnorm(0.0,1.0E-4)
 beta ~ dt(0,1,1)
 for (i in 1:length(obsx))
 {
  obsx[i] ~ dnorm(x[i],prec.obsx[i])
  obsy[i] ~ dnorm(y[i],prec.obsy[i])
  y[i] <- alpha+beta*x[i]
# t prior for the x population OR
  x[i] ~ dt(0.1,pow(1.1,-2),10)
# N(0,1) prior for the x population OR
  x[i] ~ dnorm(0,1)
# Normal prior for the x population with hyperparameters
  x[i] ~ dnorm(mu,prec)
 } 
  mu ~ dnorm(0,1.0E-4)
  prec ~ dunif(0.0,10.0)
} 
\end{verbatim}
}
\noindent
In order to predict $\tilde{y}$ values, it suffices to list the $x^{obs}$ values
for which $\tilde{y}$ is needed, entering a ``NA'' = ``not available'' code for the corresponding $y^{obs}$ value to indicate
to the program that they should be estimated.

\subsection{Code for Bayesian regression model Equation~(\ref{eqn:ex_lik2}) illustrated in Figure~\ref{fig.scatter}(a)}

{\footnotesize
\begin{verbatim}
data
{
 nu <-6
}
model 
{
 for (i in 1:length(obstot))
 {
  obsbkg[i] ~ dpois(nbkg[i])
  obstot[i] ~ dpois(nbkg[i]/C[i]+n200[i]) 
  n200[i] ~ dunif(0,3000)
  nbkg[i] ~ dunif(0,3000)
  precy[i] ~ dgamma(1.0E-5,1.0E-5)
  obslgM200[i] ~ dnorm(lgM200[i],precy[i])
  obsvarlgM200[i] ~ dgamma(0.5*nu,0.5*nu*precy[i])
  z[i] <- alpha+14.5+beta*(log(n200[i])/2.30258-1.5)
 }
 intrscat <- 1/sqrt(prec.intrscat)
 prec.intrscat ~ dgamma(1.0E-5,1.0E-5)
 alpha ~ dnorm(0.0,1.0E-4)
 beta ~ dt(0,1,1)
 }
}
\end{verbatim}
}

\bibliographystyle{unsrt}
\bibliography{draft}

{\it To be put as footnote to "references"}
Most of the mentioned papers are freely available from ADS,
http://adsabs.harvard.edu/index.html, clicking the X or G links.
\end{document}